\documentstyle[12pt,epsfig]{article}

\def\beq{ \begin{equation}}
\def\eeq{\end{equation} }
\def\bea{\begin{eqnarray}}
\def\eea{\end{eqnarray}}

\def\ohsq{\Omega_{\chi} h^2}

\def\ra{\rightarrow}

\def\bar{\overline}

\def\gappeq{\mathrel{\raise.3ex\hbox{$>$\kern-.75em\lower1ex\hbox{$\sim$}}}}
\def\lappeq{\mathrel{\raise.3ex\hbox{$<$\kern-.75em\lower1ex\hbox{$\sim$}}}}
\def    \part          {\partial}
\def    \be            {\begin{equation}}
\def    \ee            {\end{equation}}
\def    \bea           {\begin{eqnarray}}
\def    \eea           {\end{eqnarray}}
\def    \nn            {\nonumber}

\begin{document}


\title{
\begin{flushright}
\vspace*{-1.2 cm}
\vspace*{-0.4 cm}
{\normalsize FISIST/10-2002/CFIF}  \\
\end{flushright} 
\vspace*{0.7 cm} {\Large \bf Lepton-Flavour Violation in SUSY with 
and without R--parity}\footnote{To appear in the Proceedings of the Corfu
Summer Institute on Elementary Particle Physics, Corfu 2001, Edited by 
J. Rizos.}
}
\author{{\normalsize 
\bf M.~E. G\'omez and D.~F. Carvalho}\\
{\small CFIF, Departamento de Fisica, Instituto Superior T\'ecnico,
Av. Rovisco Pais,} \\
{\small 1049-001~Lisboa, Portugal} \\
\date{}}
\maketitle
\vspace*{-0.5 cm}

{\small
We study whether the individual violation of the lepton numbers
$L_{e,\mu,\tau}$ in the charged sector can lead to measurable rates
for $BR(\mu\rightarrow e \gamma)$ and $BR(\tau\rightarrow \mu \gamma)$. 
We consider three different scenarios, the fist one 
corresponds to the Minimal Supersymmetric Standard Model with 
non--universal soft terms. In the other two cases the violation of flavor 
in the leptonic charged sector is associated to the neutrino problem in models 
with a {\it see--saw} mechanism and with R--parity violation respectively.
}


\section{Introduction}

In the Standard Model (SM), lepton number is exactly preserved in
contradiction with the observed neutrino oscillations
\cite{kamiokande,chooz}. Typically, enlargements of the SM explaining 
these flavor oscillations include violation of the lepton 
numbers $L_{e,\mu,\tau}$ for charged lepton which will be 
manifested in processes such as $\mu\rightarrow e
\gamma$, $\mu\rightarrow 3 e$, $\mu-e$ conversion in heavy nuclei,
$\tau\rightarrow \mu \gamma$ and $K_L\rightarrow \mu e$. 
The experimental upper bound for these processes is quite restrictive,
which imposes a significant constraint for the explanation of flavor in
models beyond the SM. The Supersymmetric extension of the SM (SUSY) 
provides an excellent framework to study them, since the 
predicted rates can be of the order of the 
bounds that will be reached in current or projected experiments. In 
addition the SUSY contribution to the anomalous magnetic moment of 
the muon, $a_\mu\sim (g_\mu-2)/2$ can be compatible with the 
difference between the value predicted by the SM and the recent measured
value \cite{g-2}.

In this presentation we concentrate on the study of $\mu\rightarrow e
\gamma$ and $\tau\rightarrow \mu \gamma$, their current experimental 
limits are \cite{lfv-limit}:
\begin{eqnarray}
BR(\mu \to e \gamma) &<& 1.2 \times 10^{-11} \nonumber\\ 
BR(\tau \to \mu\gamma) &<& 1.1 \times 10^{-6}. 
\label{limit}
\end{eqnarray}
The SUSY predictions for these processes will be shown in three different 
scenarios. The first is  Minimal Supersymmetric SM  (MSSM) with 
non--universal soft--terms \cite{Carvalho:2000xg}, without 
involving the neutrino sector. 
The two other models can 
explain neutrino oscillations trough small neutrino masses, the 
second by means of a ''see-saw'' mechanism \cite{Carvalho:2001ex}, while in 
the third bilinear R--parity violation (BRpV) is considered \cite{paper}.

Results for other processes like $\mu-e$ conversion in heavy nuclei and 
$K_L\rightarrow \mu e$ can be found in 
refs.~\cite{Carvalho:2001ex, Ellis:1999uq, Belyaev:2000xt} and 
references there in.

\section{$l_j\rightarrow l_i \gamma$ Flavor Violating Processes and 
the $\mu$ Ano\-malous Magnetic Moment}
The effective operators that generate the decays  
$ l^{-}_{j}\rightarrow l^{-}_{i}\gamma$ and the lepton anomalous magnetic 
moment can be written as:
\begin{equation}
\label{leffg2}
L_{eff}=e\, \frac{m_{l_{j}}}{2}\, \overline{l}_{i}
\sigma _{\mu \nu } F^{\mu \nu} \left(A_{Lij}P_{L}+A_{Rij}P_{R}\right)l_{j}
\end{equation}

The one--loop contributions to  $A_{L,R}$ in R--parity conserving SUSY
arises from the diagrams in Fig. 1, while for models with BRpV 
the corresponding diagrams are presented in Fig.~4.

The branching ratio for the rare lepton decays $l_j\rightarrow l_i
\gamma$ is given by~\cite{Hisano:1995cp}:
\begin{equation}
\label{eq:brm}
BR(l^{-}_{j}\rightarrow l^{-}_{i}\gamma)=\frac{48\pi^3 \alpha }{G_F^2}
\left(\left| A_{Lij}\right| ^{2}+\left| A_{Rij}\right| ^{2}\right).
\end{equation}

The expression for the muon anomalous magnetic moment can be obtained from
the Lagrangian given in Eq.~(\ref{leffg2}):

\begin{equation}
\label{eq:amu}
a^\mu=\frac{(g_\mu-2)}{2}=-m^{2}_{\mu }(A^{\mu\mu}_{L}+A^{\mu\mu}_{R})
\end{equation}

\begin{figure}[b]
\begin{center}
\epsfig{file=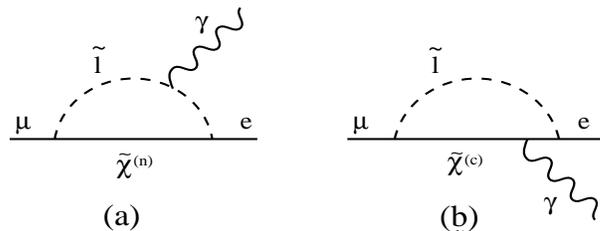,width=8cm, height=3cm}
\end{center}
\vspace*{-1cm}
\caption{{\footnotesize Generic Feynman diagrams for $\mu\ra e\gamma$
decay: $\tilde l$ represents a charged slepton (a) or
sneutrino (b), and
$\tilde\chi ^{(n)}$ and $\tilde\chi ^{(c)}$ represent  
neutralinos and charginos respectively.}}
\label{figure1}
\end{figure}

In R--parity conserving SUSY models, the presence of LFV processes is
associated with vertexes involving leptons and their
superpartners. In the Minimal Supersymmetric Standard Model
(MSSM), with  universal soft terms,  
it is possible to rotate the charged lepton Yukawa couplings
and the sleptons in such a way that lepton flavor is preserved. However,
a small deviation from flavor universality in the soft-terms at the GUT
scale will be severely constrained by the experimental
bounds~(\ref{limit}). In fact GUT theories ~\cite{rh-nu}
and models with $U(1)$~family symmetries~\cite{Gomez:1998wj}
can lead to the  MSSM with flavor--dependent soft terms leading to
important violations of the lepton flavor.

The lepton Yukawa couplings can be diagonalized by the
unitary matrices $U_L$ and $U_R$ as follows,
\be
m_l = \frac{v \cos
\beta}{\sqrt{2}} U_R~ (Y^l)^T~ U^{\dag}_L .
\ee
When the superfields are written in this basis, the 
expressions for the charged
slepton mass matrices at low energy take the form:
\bea
\begin{array}{lr}
M^2_{\tilde{l}}=
\left(\begin{array}{cc}
\left(M^2_{\tilde{l}}\right)_{LL}
&\left(M^2_{\tilde{l}}\right)_{LR}
\\
\left(M^2_{\tilde{l}}\right)_{RL}
&\left(M^2_{\tilde{l}}\right)_{RR}
\end{array} \right),
\end{array}
\eea
where,
\bea
\left(M^2_{\tilde{l}}\right)_{LL}&=&U_L m^2_{L} U_L^{\dag}
+m_l^2 -\frac{m_Z^2}{2}(1-2\sin^2{\theta_W})\cos{2\beta},\nn \\
\left(M^2_{\tilde{l}}\right)_{RR}&=&U_R (m^2_{R})^T U_R^{\dag}
+m_l^2  + m_Z^2 \sin^2{\theta_W}\cos{2\beta},\nn \\
\left(M^2_{\tilde{l}}\right)_{LR}&=&\left(M^2_{\tilde{l}}\right)_{RL}^{\dag}=
-\mu~m_l \tan{\beta}+\frac{v\cos{\beta}}{\sqrt{2}}
U_L Y^{A *}_l U_R^{\dag} ,
\eea
where $m^2_{L}$ and $m^2_{R}$ are the soft breaking $(3\times3)$ mass
matrices for the slepton doublet and singlet respectively.

The sneutrino mass matrix is simply  given  by the $(3\times3)$ mass matrix:
\be
M^2_{\tilde{\nu}}=U_L m^2_{L} U_L^{\dag}+
\frac{m_Z^2}{2}\cos{2\beta}
\label{sneutrino}
\ee
The relevant lepton--flavor changing mass matrix elements
on the slepton  mass matrices above are given by:

\bea
(\delta^l_{LL})_{ij} &=& \left[~ U_L~ m^2_{L}
~U_L^{\dag}~ \right]_{ij}\nn \\
(\delta^l_{LR})_{ij} &=&  \left[~ U_L~ Y^{A *}_l
~U_R^{\dag}~ \right]_{ij} \label{Deltas} \\
(\delta^l_{RR})_{ij} &=& \left[~ U_R~
(m^2_{R})^T~ U_R^{\dag}~ \right]_{ij} \nn
\eea
where $i$, $j$ are flavor indices ( $i\neq j$).

\section{Models with Non-universal Soft SUSY breaking terms}
The example of the MSSM with non--universal soft
SUSY breaking terms presented here is based on the effective 
supergravity theories  which emerge in the low energy limit of the 
weakly coupled heterotic strings (WCHS)~\cite{ibanez1}.

In the WCHS, it is assumed that the superpotential of
the dilaton ($S$) and moduli ($T$) fields is
generated by some non--perturbative mechanism, and that
the $F$-terms of $S$ and $T$ contribute to the SUSY breaking.
Hence one can parametrize the $F$-terms as
\be
F^S = \sqrt{3} m_{3/2} (S+S^*) \sin\theta,\hspace{0.75cm} F^T
=m_{3/2} (T+T^*) \cos\theta .
\ee
Here $m_{3/2}$ is the gravitino mass and $\tan \theta$ corresponds to the 
ratio between the $F$-terms of $S$ and $T$.
In this framework, the soft scalar masses $m_i$ and the gaugino masses
$M_a$ are given by
\begin{eqnarray}
m^2_i &=& m^2_{3/2}(1 + n_i \cos^2\theta), \label{scalar}\\ M_a
&=& \sqrt{3} m_{3/2} \sin\theta ,\label{gaugino}
\end{eqnarray}
where $n_i$ is the modular weight of the corresponding field. The $A$-terms are written as
\begin{eqnarray}
A_{ijk} &=& - \sqrt{3} m_{3/2} \sin\theta- m_{3/2}
\cos\theta (3 + n_i + n_j + n_{k}), \label{trilinear}
\end{eqnarray}
where $n_{i,j,k}$ are the modular weights of the fields
that are coupled by this $A$--term. We assume $n_i=-2$ for the first two 
generations and for the down Higgs and $n_i=-1$ for the third generation and 
the up Higgs.

To illustrate the dependence of the results on the lepton Yukawa 
couplings, we consider two examples of symmetric textures at the GUT scale:

\begin{itemize}
\item Texture I,
\be
Y_{l}=y^{\tau }\left( \begin{array}{ccc}
0 & 5.07\times 10^{-3} & 0\\
5.07\times 10^{-3} & 8.37\times 10^{-2} & 0.4\\
0 & 0.4 & 1
\end{array}\right)
\ee 
\item Texture II,
\be
Y_{l}=y^{\tau }\left( \begin{array}{ccc}
3.3\times 10^{-4} & 1.64\times 10^{-5} & 0\\
1.64\times 10^{-5} & 8.55\times 10^{-2} & 0.4\\
0 & 0.4 & 1
\end{array}\right)
\ee
\end{itemize}

Texture I is a symmetric texture which can be considered to be the 
limiting case of textures arising from  $U(1)$~family 
symmetries as described in Refs.~\cite{u1m} and studied in the next section.
Typically a prediction for the decay
 $\tau \to \mu \gamma$ of the order of the experimental limit~(\ref{limit})
will imply a severe violation of the experimental bound for
 $\mu \to e \gamma$. We can see in Fig.~2 how texture I (graphic on the left) 
tolerates small deviations from universality of the soft terms. 
The experimental bound on $BR(\mu\to e \gamma)$ is satisfied only 
for $\sin \theta> .96$ ($m_{3/2}=200\ \rm{GeV}$) and for $\sin \theta> .91$ 
($m_{3/2}=400\ \rm{GeV}$) while for the same range on  $\sin \theta$
the corresponding prediction for $BR(\tau\to \mu \gamma)$ is well below
the experimental bound.

Texture II was chosen as an 
illustration of how the current bounds (\ref{limit}) can provide 
some information about the lepton Yukawa couplings on the context of the 
models considered. The results obtained using texture II 
(Fig.~2, graphic on the right) allow us to start the graph at the 
lowest value of $\sin\theta=1/\sqrt{2}$. As it can  be seen, the 
experimental bounds are more restrictive for the $\tau \to \mu \gamma$ 
than for $\mu \to e\gamma$ process.


\begin{figure}[t]
\hspace*{-0.5in}
\begin{center}
\begin{minipage}{7in}
\epsfig{figure=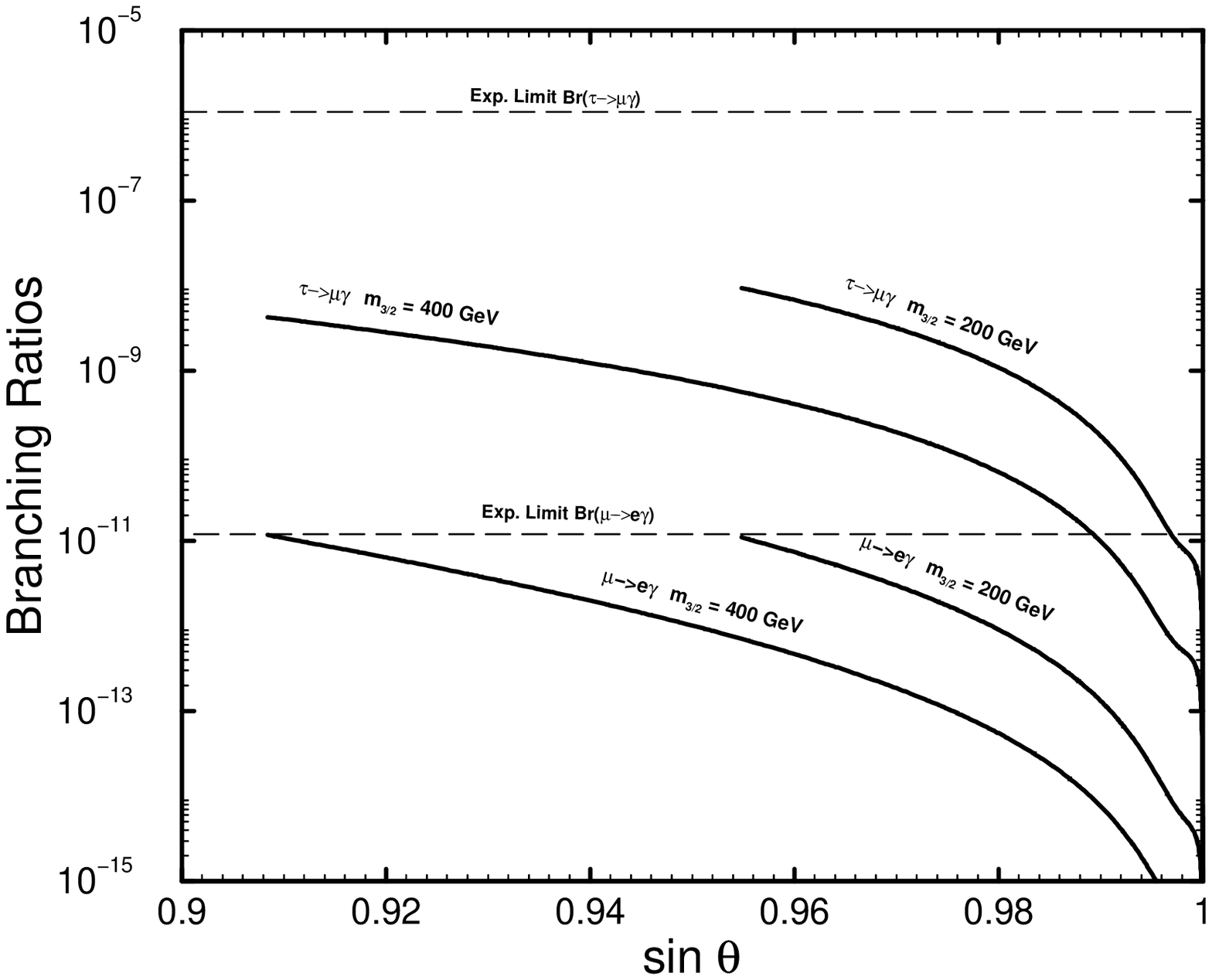,height=2.8in,width=2.8in,angle=0}
\epsfig{figure=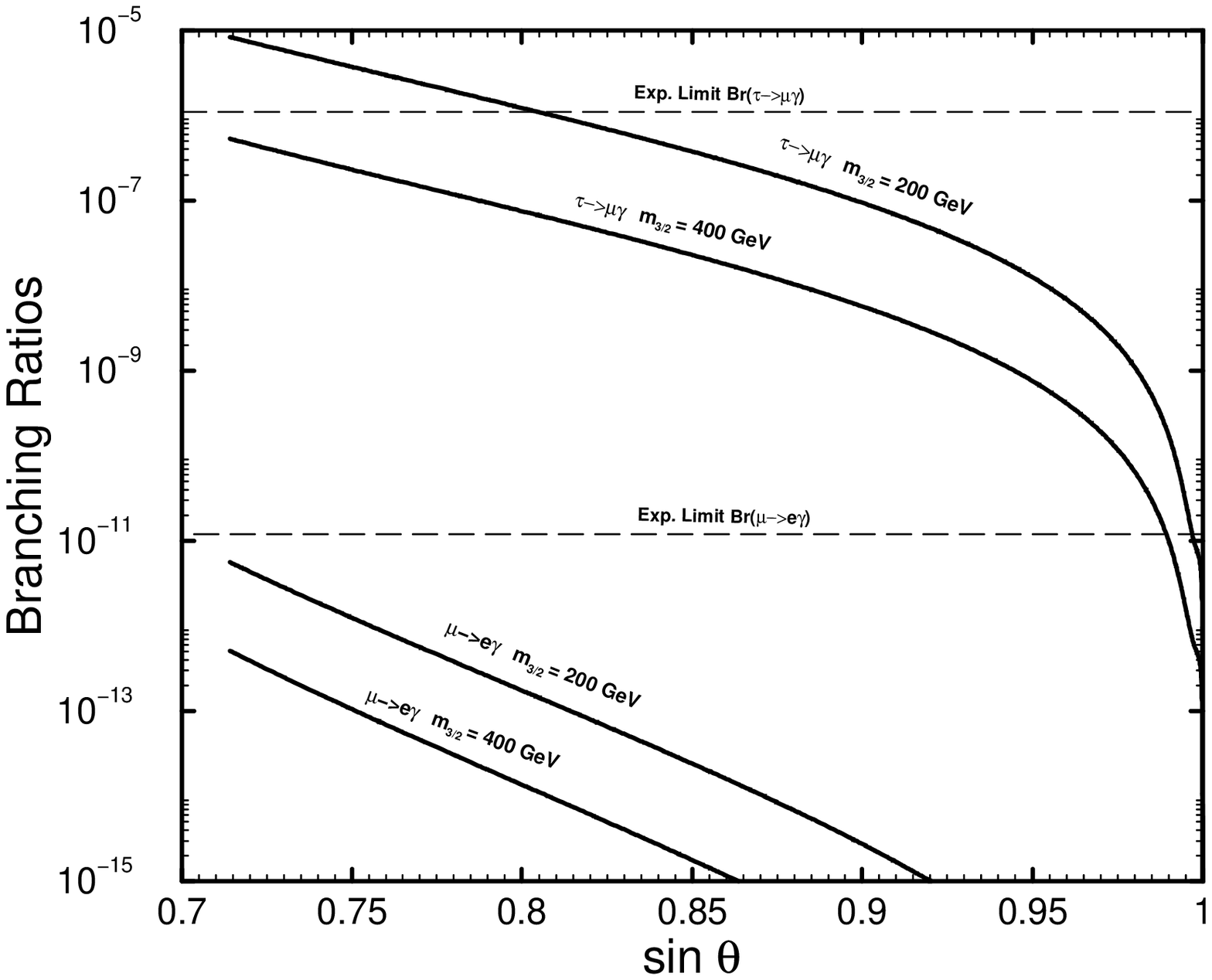,height=2.8in,width=2.8in,angle=0}
\end{minipage}
\end{center}
\vspace*{-1cm}
\caption{{\footnotesize Branching ratios vs. $\sin\theta$ for the WCHS model with
texture I for $Y_l$ (left) and texture II (right) and $\tan\beta=10$. 
The values for $m_{3/2}$ are kept constant as shown on the curves.}
\label{fig.2}}
\end{figure}

\section{SUSY Models with {\bf{\it''see--saw''}} Mechanism}

One of most attractive mechanisms for obtaining sub-eV neutrino masses is 
the {\it''see--saw''} mechanism~\cite{seesaw}. The ''see-saw'' mechanism 
can be 
included in SUSY models by enlarging the MSSM with right-handed neutral 
fields, $N^c$, such that the superpotential at the GUT scale becomes:
\begin{equation}
W=W_{MSSM}+N^c h_D L H_u+ \lambda_N \chi N^c N^c;
\end{equation} 

Were $\chi$ is a singlet field which acquires a VEV at a high scale of 
energy leading to heavy Majorana masses for the fields $N^c$, 
$M_N= \lambda_N <\chi>$.

The effective theory below the $M_N$ \cite{Hisano:1995cp,Casas:2001sr} scale 
(which here is assumed common for the three generations) is the the 
MSSM with tiny neutrino masses:

\begin{equation}
W_{eff}= W_{MSSM} +(h_D L H_u)^T M_N^{-1} (h_D L H_u);
\end{equation}

\begin{figure}[t]
\begin{center}
\epsfig{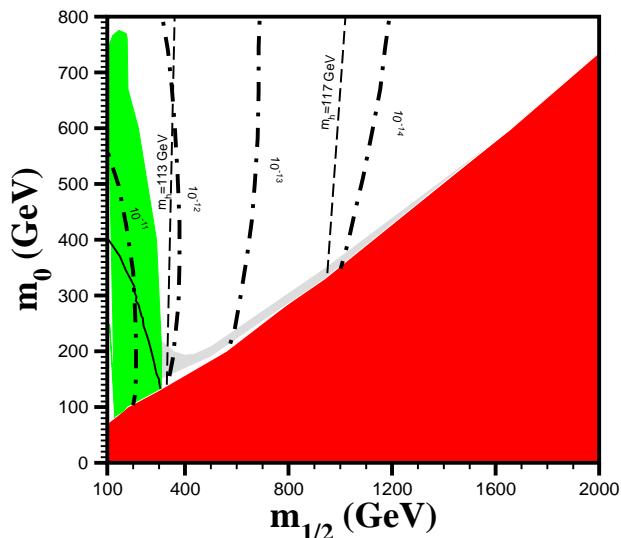}
\end{center}
\vspace*{-1cm}
\caption{\label{fig:muegamma}
{\footnotesize The contours BR$(\mu \ra e \gamma) = 10^{-11}, 10^{-12},
10^{-13}$ and $10^{-14}$ are shown as
dash-dotted black lines in the $(m_{1/2}, m_0)$ planes for $\mu > 0$ and
$\tan \beta =$ 30. Other constraints in these planes
are taken from~\cite{Ellis:2001yu}, assuming $A_0 = 0, m_t = 175$~GeV and
$m_b(m_b)^{\overline {MS}}_{SM} = 4.25$~GeV. The region allowed by the
E821 measurement of $a_\mu$ at the 2-$\sigma$ level is the area on the 
right of the solid black line. The dark (red) shaded regions are excluded
because the LSP is the charged ${\tilde \tau}_1$, and the light
(gray) shaded regions are those with \protect\mbox{$0.1\leq\ohsq\leq
0.3$} that are preferred by cosmology. We show the contours $m_h = 113,
117$~GeV.  The medium (green) shaded regions are 
excluded by $b \to s \gamma$.  }}
\end{figure}

In general, the Dirac neutrino and charged-lepton Yukawa
couplings,  $h_{D}$ and $Y_{l}$ respectively,
cannot be diagonalized simultaneously.
Since both these sets of lepton Yukawa
couplings appear 
in the renormalization-group equations,
the lepton Yukawa
matrices and the slepton mass matrices cannot be 
diagonalized simultaneously at low energies, either.
In the basis where $Y_{l}$ is diagonal, the slepton-mass matrix acquires 
non-diagonal contributions from renormalization at scales below
$M_{GUT}$ \cite{Casas:2001sr}, of the form:
\bea
(\delta_{LL}^l)\propto \frac 1{16\pi^2} (3 + a^2)
\ln\frac{M_{GUT}}{M_N}h_{D}^{\dagger} h_{D} m_{3/2}^2,
\label{offdiagonal}
\eea
where $a$ is  related to the trilinear mass parameter:
$A_\ell \equiv a m_{0}$, where
$m_0$ is the common assumed value of the scalar masses at the GUT
scale.

In order to illustrate our estimates of the expected effects, we calculate
the rates for rare processes violating charged-lepton number in
a model with Abelian favor symmetries and 
symmetric fermion mass matrices~\cite{u1m}, which leads to the following 
pattern of charged-lepton masses $m_\ell$ and neutrino Dirac 
masses $m_{\nu_D}$:
\bea
m_{\ell }  \propto  \left( 
\begin{array}{ccc}
\bar{\epsilon}^{7} & \bar{\epsilon}^{3} & \bar{\epsilon}^{7/2} \\ 
\bar{\epsilon}^{3} & \bar{\epsilon} & \bar{\epsilon}^{1/2} \\ 
\bar{\epsilon}^{7/2} & \bar{\epsilon}^{1/2} & 1
\end{array}
\right),
m_{\nu_D} \propto \left( 
\begin{array}{ccc}
\bar{\epsilon}^{14} & \bar{\epsilon}^{6} & \bar{\epsilon}^{7} \\ 
\bar{\epsilon}^{6} & \bar{\epsilon}^2 & \bar{\epsilon} \\ 
\bar{\epsilon}^{7} & \bar{\epsilon} & 1
\end{array}
\right),
\eea

where $\bar{\epsilon}$ is a (small) expansion parameter
related to the Abelian symmetry-breaking scale 
(assumed here to be $\approx .2)$.

The elements of the Yukawa coupling matrices at the GUT 
scale can be chosen to be
consistent with the experimental values of the fermion masses 
by introducing  coefficients of order one  in the entries of the mass 
matrices. In the notation of~\cite{Ellis:1999uq}, we choose for
this model (model A ) $C_{12} \, = \, 0.77, \; \; C_{23} \, = \, 0.79$.

The difference on the value of the muon anomalous magnetic moment
found in the BNL E821 measurement \cite{g-2} with respect to
the SM prediction, which originally was considered to be $2.6\,
\sigma$ is now reduced to $1.6\, \sigma$ after a theoretical error has
been corrected (see discussion and references in \cite{paper}). When 
the $2\, \sigma$ range is considered,
the allowed values for contributions beyond the SM become,
\begin{equation}
\label{eq:g2range}
-6\times 10^{-10}\leq \delta a_\mu \equiv  
a_\mu^{exp}-a_\mu^{SM}\leq 58\times 10^{-10},
\end{equation}

The predictions for BR($\mu \ra e \gamma$) are shown in Fig.~3, the 
solid line correspond to the upper bound of (\ref{eq:g2range}). The 
other constraints described in the caption are taken from \cite{Ellis:2001yu}. 
The predictions for B($\tau \rightarrow  \mu \gamma$) in this model 
can be found in \cite{Ellis:1999uq}, these are typically less restrictive 
than the case of $\mu\ra e \gamma$ discussed above.

\section{SUSY with Bilinear R--Parity Violation} 
The simplest extension of the MSSM with bilinear R--parity violation
(BRpV) (allowing B--conserving but L-violating
interactions) can explain the neutrino masses and mixings which can
account for the observed neutrino oscillations \cite{Romao:1999up}.  

In the model we consider, the MSSM superpotential is enlarged with 
bilinear terms that violate lepton number and therefore also
breaks R--parity:
\begin{equation}
\label{eq:Wsuppot}
W = W_{MSSM} + \epsilon_{i}{L}_{i} {H}_{u}.
\end{equation}

The inclusion of the R--parity violating term, though
small, can modify the predictions of the MSSM. The most salient
features are that neutrinos become massive and the lightest neutralino
is no longer a dark matter candidate because it is 
allowed to decay. Furthermore, we
can observe that this model implies the mixing of the leptons with the
usual charginos and neutralinos of the MSSM. Lepton Yukawa couplings can be
written as diagonal matrices without any loss of generality since it
is possible to rotate the superfields \( {L}_{i} \) in the
superpotential, Eq.~(\ref{eq:Wsuppot}), such that Yukawa matrix \(
Y_{l} \) becomes diagonal. Conversely, in BRpV models it is possible
to apply a similar rotation to reduce the number \( \epsilon \)
parameters and provide a non-trivial structure to \( {Y}_{l} \)
\cite{Gomez:1998sd}.

 In addition to the MSSM soft SUSY breaking terms in \(
V_{soft}^{MSSM} \) the BRpV model contains the following
extra term
\begin{equation}
\label{softBRpV}
V_{soft}^{BRpV}=-B_{i}\epsilon _{i} \widetilde{L}_{i} H_{u}.
\end{equation}

The electroweak symmetry is broken when the two Higgs doublets \( H_{d} \)
and \( H_{u} \), and the neutral component of the slepton doublets 
\( \widetilde{L}_{i} \) acquire vacuum expectation values.

In addition to the above MSSM parameters, our model
contains nine new parameters, \( \epsilon _{i} \), \( v_{i} \) and \( B_{i} \).
The minimization of the scalar potential allows to relate some of these free
parameters.

The range of values of the $\epsilon$--parameters is indirectly
associated to the size of the neutrino masses predicted by the
model. To explore this relation we describe next the mass mixings
among neutralinos and neutrinos.  In the basis \( \psi ^{0T}=
(-i\lambda' ,-i\lambda ^{3},\widetilde{H}_{d}^{0},
\widetilde{H}_{u}^{0},\nu _{e},\nu _{\mu },\nu _{\tau }) \) the
neutral fermion mass matrix \( {\bf M}_{N} \) is given by
\beq
{\bf M}_N=\left[ \begin{array}{cc}  
{\mathcal{M}}_{\chi^{0}}& m^{T} \cr
m & 0 \cr
\end{array} \right] 
\eeq
where ${\mathcal{M}}_{\chi^{0}}$  is the standard MSSM neutralino 
mass matrix and 
\beq
m=\left[ \begin{array}{cccc}  
-\frac {1}{2}g^{\prime }v_{1} & \frac {1}{2}gv_{1} & 0 & \epsilon _{1} \cr
-\frac {1}{2}g^{\prime }v_{2} & \frac {1}{2}gv_{2} & 0 & \epsilon _{2}  \cr
-\frac {1}{2}g^{\prime }v_{3} & \frac {1}{2}gv_{3} & 0 & \epsilon _{3}  \cr  
\end{array} \right] 
\eeq
characterizes the breaking of R-parity. 

The mass matrix \( {\bf M}_{N} \) is diagonalized by
\beq
{\mathcal{N}}^{*}\bf M_N{\mathcal{N}}^{-1}=diag(m_{\chi^{0}_i},
m_{\nu_j}),  
\label{chi0massmat} 
\eeq
where \( (i=1,\cdots ,4) \) for the neutralinos, and \( (j=1,\cdots ,3) \)
for the neutrinos.
One of the neutrino species acquire a tree level non--zero mass, given by:
\beq
\label{mnutree} m_{\nu_{3}} = Tr(m_{eff}) = 
\frac{M_{1} g^{2} + M_{2} {g'}^{2}}{4\, \det({\mathcal{M}}_{\chi^{0}})}
|\vec {\Lambda}|^{2}, 
\eeq
where $|\vec {\Lambda}|^{2}\equiv \sum_{i=1}^3 \Lambda_i^2$. The 
$ \Lambda_i$ parameters are defined as: 
\beq
\Lambda_i \equiv \mu v_i + v_d \epsilon_i
\label{lambdai} 
\eeq
The two other neutrinos can get masses at one--loop as 
it is discussed in Ref.~\cite{Romao:1999up}. For our purposes it will
be important to have an estimate of the values of $\Delta m^2_{12}=
m^2_{\nu_2}- m^2_{\nu_1}$. We will use the results of
Ref.~\cite{Corfu} where it was found that, to a very good
approximation, $m_{\nu_1}=0$ and 
\begin{equation}
  \label{eq:mneu2}
  m_{\nu_2}=\frac{3}{16\pi^2}\, m_b \, \sin^ 2 \theta_b
  \,\frac{h^2_b}{\mu^2}\, \log \frac{m^2_{\tilde{b}_2}}{m^2_{\tilde{b}_1}}\
  \frac{\left( \vec \epsilon \times \vec \Lambda \right)^2 }{|\vec
  \Lambda |^2}
\end{equation}

The one--loop contributions to  $A_{L,R}$ in eq.~\ref{leffg2} arise from 
the diagrams of Fig.~4. We
follow the notation of \cite{paper, Romao:1999up} indicating by $S^{\pm }$ the
eigenstates of the charged scalar mass matrix, by $S^{0}$ and $P^{0}$
the eigenstates for the sneutrino--Higgs scalar mass matrices,
CP--even and CP--odd, respectively.

\begin{figure}[t]
\begin{center}
\vspace*{-.7cm}
\epsfig{file=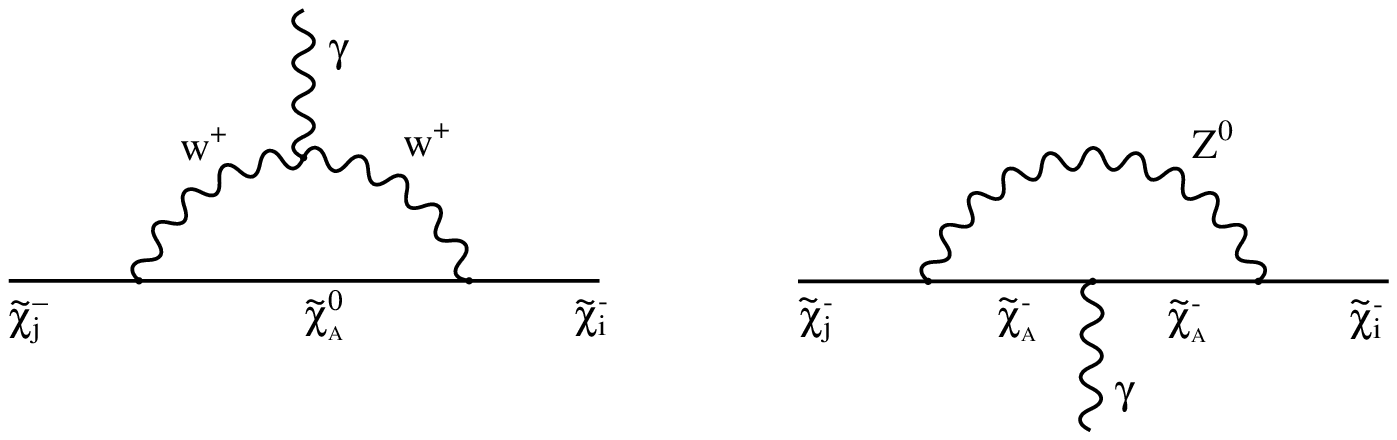,width=8cm,height=3cm}
\vspace*{-.7cm}
\epsfig{file=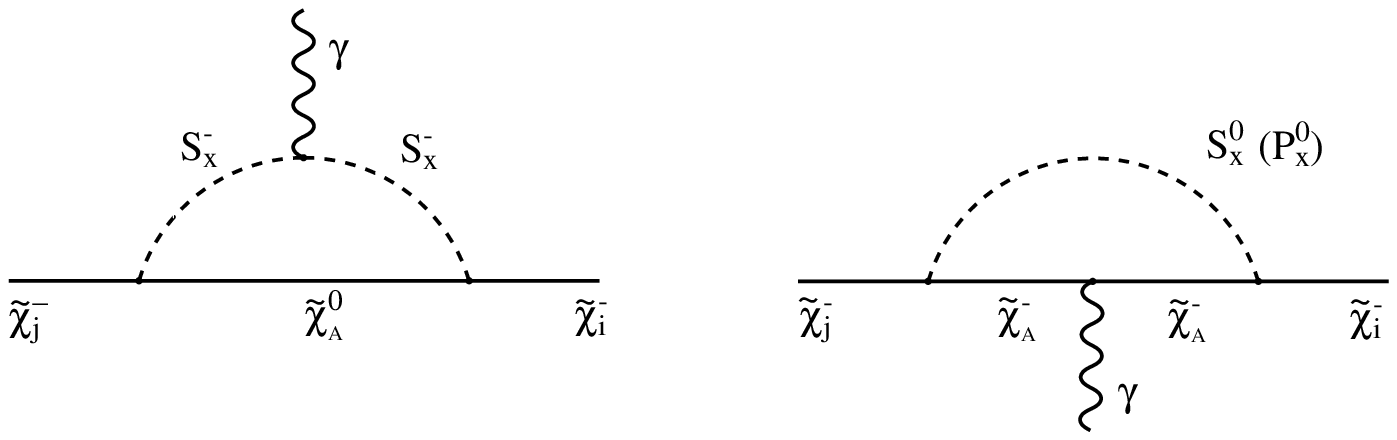,width=8cm,height=3cm}
\vspace*{-.7cm}
\epsfig{file=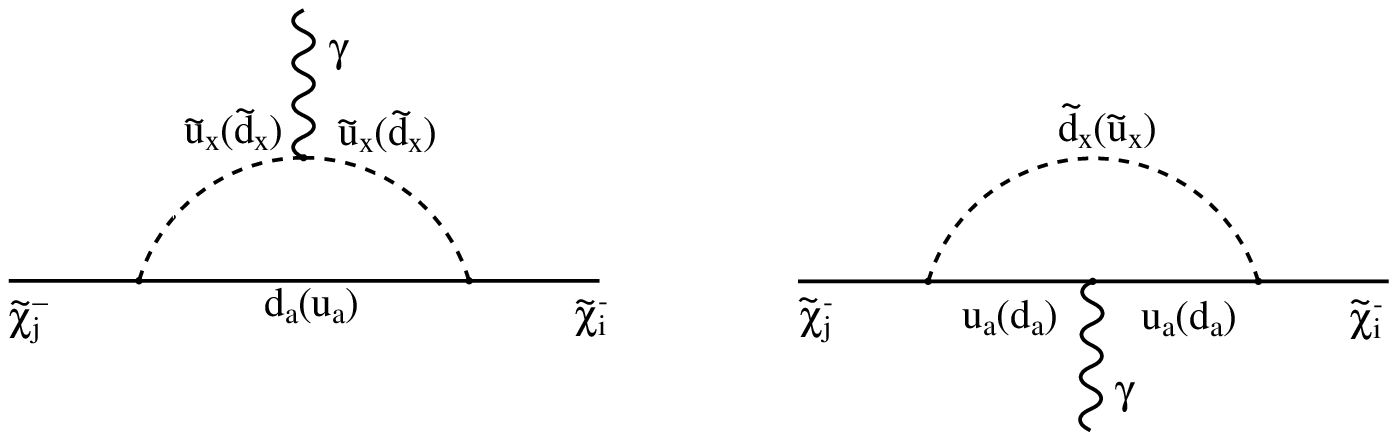,width=8cm,height=3cm}
\end{center}
\vspace{-10mm}
\caption{\footnotesize Generic Feynman diagrams 
for $A_{L/R}$ in the BRpV model.}
\label{fig2}
\end{figure}

For the example considered in Fig.~5 we take, 
$\tan\beta=30$, $m_{1/2}=400\ {\rm GeV}$, $m_0=300\ {\rm GeV}$, 
$A_0=0$, in order to compare our results with the ones presented in 
the previous section. We take a $m_{\nu_3} = 0.1\ \rm{eV}$, which
leads to values of the $|\vec {\Lambda}|$ in the range of $0.1-1\
\rm{GeV}^2$, for the values of the SUSY parameters that we will
consider. Considering that we take positive values for $\mu$ we should 
also take negative values for the product $\epsilon_i v_i$ 
to avoid our analysis to be constrained to small values of $\epsilon_i$.

The six free BRpV breaking parameters \( \epsilon _{i},\, \, v_{i} \)
reduce to three if we take into account the constraints imposed by the
predictions for neutrino oscillations in this model, as given in
Ref.~\cite{Romao:1999up}. It was shown in this reference that the 
conditions \(\Lambda_{3}\simeq\Lambda_{2}\simeq 5\times \Lambda_{1} \) 
satisfy both the atmospheric neutrino anomaly mixings and the CHOOZ
result~\cite{chooz}. We then obtain a linear relationship between each
couple \( \epsilon _{i},\, \, v_{i} \). 

We can compare the results presented in Fig.~5 with predictions  
for $BR(\mu \rightarrow e \gamma)$  presented in the previous section. 
As we can see in Fig.~3, the choice of MSSM parameters used in Fig.~5
corresponds to a value between $10^{-12}$
and $10^{-13}$ for $BR(\mu \rightarrow e \gamma)$.
These values will be reached in the BRpV case for values of $|\epsilon_1|$
and $|\epsilon_2|$ ranging from $1$ to $10$~GeV (independently of the
value of $\epsilon_3$). Values in the range of $0.1$ to $1$~ GeV would 
lead to rates of order $10^{-14}-10^{-16}$, still interesting for the 
next generation experiments \cite{psi,prism}. Such values of $|\epsilon_i|$ 
are however excluded if one takes into account the constraint coming 
from the solar neutrinos mass scale. This is 
shown in Fig.~5,  where the dashed line gives the upper limit on 
the $|\epsilon_i|$ as obtained from Eq.~(\ref{eq:mneu2}) for the 
requirement that $m_{\nu_2} < 0.01\ \rm{eV}$. 
As it is discussed in Ref.~\cite{paper}, the parameters which enhances 
the ratios also make $m_{\nu_2}$ larger. From eq.(\ref{eq:mneu2}) we can 
observe that $m_{\nu_2}$ increases with $\tan\beta$ (through the 
dependence on $h_b$) and decreases as the $\mu$-term increases 
(i.e with $m_{1/2}$ and $m_0$). 

Contributions to $\delta a_\mu$  arising from the BRpV terms are found to be 
small compared with the MSSM limit. The corresponding values for the 
partial contributions from the diagrams of Fig.~4 can be found in 
Ref.~\cite{paper}. Also the prediction of the model 
for $ \tau \rightarrow \mu \gamma$ rates is of the same order of 
the $BR(\mu \rightarrow e \gamma)$ presented here, therefore out of the 
experimental range. 

\section{Conclusions}
We had reviewed the predictions for the rare lepton decays 
BR($ \mu \rightarrow e\gamma$) and BR($ \tau \rightarrow \mu \gamma$) in the 
context of SUSY models. The first scenario was used to explain how the 
nature of the soft--terms combined with a non--trivial texture for the 
charged lepton Yukawa couplings turns into a prediction for charged lepton 
flavor violation. In the second scenario, these conditions are obtained when 
neutrino flavor oscillations are explained trough a 
{\it ``see-saw''} mechanism. The predicted rates in both scenarios 
are of experimental interest and will be tested at  PSI~\cite{psi} or at
PRISM~\cite{prism} providing a relevant information on the free parameters 
of the models.

The obtained results for the $\mu
\rightarrow e \gamma$ in our third scenario show us that if the BRpV 
model is the explanation for both the solar and atmospheric neutrino 
oscillations, the predicted LFV will not be testable in planned experiments. 
The correlations of the BRpV parameters with the
neutralino decays, as proposed in Ref.~\cite{Porod:2000hv}, will
remain the main test of the model. 

\begin{figure}[t]
\vspace*{-1cm}
    \begin{center}
      \epsfig{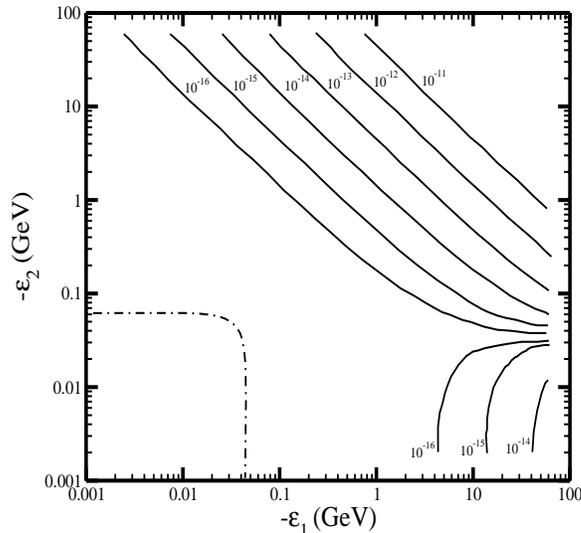}
    \end{center}
\vspace*{-1cm}
    \caption{\footnotesize Contour plot for BR($ \mu \rightarrow e\gamma$) in 
            $ \epsilon _{1}$--$\epsilon _{2}$ plane. The dash lines 
correspond to $m_{\nu_2}=.01$~eV.}
    \label{fig6}
\end{figure}

{\bf ACKNOWLEDGMENTS:} This project was supported in part 
by the TMR Network of the EC under
contract HPRN-CT-2000-00148. We acknowledge
support from the `Funda\c c\~ao para a Ci\^encia e Tecnologia' 
under contracts PRAXIS XXI/
BD/9416/96 (D. F. C.) and SFRH/BPD/5711/2001 (M. E. G).

\end{document}